\documentclass[useAMS,espf]{mn2e}
\usepackage{epsfig}
\usepackage{rotating}
\title[Strong X-ray variability in NGC~1851]{Large amplitude
variability from the persistent ultracompact X-ray binary in NGC~1851}
\author[T.J Maccarone et al.]{Thomas J. Maccarone \\ School of Physics
and Astronomy, University of Southampton, Hampshire SO17 1BJ,United
Kingdom\\ \newauthor Knox S. Long\\ Space Telescope Science Institute,
Baltimore MD 21218, USA\\ \newauthor Christian Knigge, Andrea
Dieball\\ School of Physics and Astronomy, University of Southampton,
Hampshire SO17 1BJ,United Kingdom\\ \newauthor David R. Zurek\\
Department of Astrophysics, American Museum of Natural History, New
York NY 10024, USA\\ }

\begin{document}
\def\gtsim{\mathrel{\rlap{\lower 3pt\hbox{$\sim$}}
        \raise 2.0pt\hbox{$>$}}}

\date{}

\pagerange{\pageref{firstpage}--\pageref{lastpage}} \pubyear{}

\maketitle

\label{firstpage}

\begin{abstract}
Using archival RXTE data, we show that the ultracompact X-ray binary
in NGC~1851 exhibits large amplitude X-ray flux varations of more than
a factor of 10 on timescales of days to weeks and undergoes sustained
periods of months where the time-averaged luminosty varies by factors
of two.  Variations of this magnitude and timescale have not been
reported previously in other ultracompact X-ray binaries.  Mass
transfer in ultracompact binaries is thought to be driven by
gravitational radiation and the predicted transfer rates are so high
that the disks of ultracompact binaries with orbits as short as that
of this object should not be susceptible to ionization instabilities.
Therefore the variability characteristics we observe were unexpected,
and need to be understood.  We briefly discuss a few alternatives for
producing the observed variations in light of the fact that the
viscous timescale of the disk is of order a week, comparable to the
shorter time scale variation that is observed but much less than the
longer term variation.  We also discuss the implications for
interpretation of observations of extragalactic binaries if the type
of variability seen in the source in NGC~1851 is typical.

\end{abstract}

\begin{keywords}
accretion, accretion discs -- X-rays:binaries -- X-rays:individual:4U~0513-40 -- globular clusters: individual: NGC~1851
\end{keywords}

\section{Introduction}
Ultracompact X-ray binaries are semi-detached binary star systems in
which a neutron star or a black hole accretes material from a
Roche-lobe overflowing white dwarf.  They are of astrophysical
interest for a variety of reasons.  It is expected that some
ultracompact X-ray binaries should be detectable as gravitational wave
sources with missions like LISA (see e.g. Benacquisita 1999).
Additionally, they present an opportunity to study plasma astrophysics
in hydrogen-free gas, where the mass-to-charge ratio is different than
in most other astrophysical systems.

The most recent compilation of ultracompact X-ray binaries in the
Galaxy contained 27 candidates (in 't Zand, Jonker \& Markwardt 2007
-- IZJM07).  The most secure identifications of ultracompact X-ray
binaries come through direct measurements of their orbital periods,
ranging in the observed sample from 11 to 50 minutes (IZJM; Galloway
et al. 2002; Middleditch et al. 1981; Markwardt et al. 2002, 2003;
Stella et al. 1987; Homer et al. 1996; White \& Swank 1982; Dieball et
al. 2005; and Zurek et al. 2009 for the period measurement of the
object discussed in this paper).  At the time of publication of
IZJM07, only 7 had well-estimated orbital periods.  The remainder were
classified as ultracompact X-ray binaries on the bases of some
combination of tentative orbital period measurements, deep optical
spectra lacking hydrogen emission lines, high ratios of X-ray to
optical flux, or persistent emission at low fractions of the Eddington
rate (IZJM07 and references within).  The persistent ultracompact
X-ray binaries typically have X-ray luminosities from about
$10^{36}-10^{37}$ ergs/sec.

Understanding the formation mechanisms for ultracompact X-ray binaries
is also of great interest.  The shortest period ultracompact X-ray
binaries (i.e. those with orbital periods less than 30 minutes), for
example, are predominantly in globular clusters (see e.g. the
tabulation in IZJM07), and thus their production may be dominated by
mechanisms which are inherently stellar dynamical, such as direct
collisions between neutron stars and red giants (e.g. Verbunt 1987).
However, theoretical work does suggest that it is possible to form
white dwarf-neutron star binaries with orbital periods less than 30
minutes by going through an intermediate phase where the donor star is
a helium star (e.g. Savonije, de Kool \& van den Heuvel 1986).  At
least the shortest period ultracompact binaries are relatively easy to
detect, as they are bright ($L_X \gtsim 10^{36}$ ergs/sec), persistent
X-ray emitters. These thus represent samples of double degenerate
stars with relatively well understood selection effects, and which can
be seen nearly throughout the Galaxy.  Other classes of binary compact
objects which are of great astrophysical interest include double white
dwarfs, which may be the progenitors of Type Ia supernovae (Iben \&
Tutukov 1984) and double neutron stars, which are showing increasingly
strong evidence for being the progenitors of short-hard gamma-ray
bursts (e.g. Bloom et al. 2006).

Understanding the evolution of double compact binaries is likely to
require using multiple source classes to gather constraints.
Double neutron stars, for example, are both rare, and very difficult
to detect.  Mass transferring double white dwarf systems can be
studied in great deal once they are detected (e.g. Roelofs et
al. 2007), but their sample sizes are only slightly larger than those
for the ultracompact X-ray binaries, and the relative robustness to
selection effects in surveys of neutron star ultracompact X-ray
binaries and mass transferring double white dwarf systems is not yet
clear.

In the absence of orbital period estimates, candidate ultracompact
X-ray binaries can be found through measurements of large X-ray to
optical flux ratios (e.g. Juett et al. 2001).  This signature is
expected for short orbital period systems since the optical emission
from most bright X-ray binaries is dominated by reprocessing of X-rays
in the outer accretion disk, leading to a correlation between X-ray to
optical flux ratio and orbital period (van Paradijs \& McClintock
1994).  Additional candidate UCXBs can be identified from strong upper
limits on hydrogen emission lines in optical spectra (e.g. Nelemans et
al. 2004), or by detection of persistent bright hard X-ray emission
combined with low rates of Type I X-ray bursts (IZJM07).

Bildsten \& Deloye (2004) have suggested that ultracompact X-ray
binaries may be the dominant population in elliptical galaxies.  The
typical X-ray luminosity functions seen in elliptical galaxies
(e.g. Kundu et al. 2002; Kim \& Fabbiano 2004; Gilfanov 2004), and the
similarities between the luminosity functions of globular cluster
X-ray sources and non-cluster X-ray sources in the elliptical galaxies
(Kundu et al. 2002) are a major part of the evidence they suggest for
this idea.  The elliptical galaxy X-ray observations used in this
analysis typically consist of snapshots of about half a day.  Strong
variability can cause the observed luminosity functions in snapshots
to be different from the luminosity function which would be obtained
when averaging over long durations.

In this paper, we present the X-ray light curve of 4U~0513-40 from
RXTE pointed observations.  This X-ray binary, in the Galactic
globular cluster NGC~1851, has a 17-minute orbital period -- short
enough that irradiated disc models (Lasota et al. 2008) predict it
should not be subject to the standard ionisation instability (i.e. it
should always be sufficiently ionised to exist in the high viscosity
state), and hence should persistently be a bright X-ray source.  It
{\it is} persistent in the sense that it is always detectable above
$L_X$ of about $10^{36}$ ergs/second.  However, we also present
evidence for variability of a factor of $\sim10$ in X-ray luminosity
on timescales of $\sim$ weeks, and a factor of more than 20 overall.
Such variability is unusual for ultracompact X-ray binaries, and
contrary to the expectations from ionisation instability theory for
such source.  4U~0513-40 also shows variations at the factor of 2
level in the time averaged luminosity over years, with no strong
indications of any periodicities in the long term light curve.

\section{Data used, analysis procedure, and results}

We have analyzed the archival RXTE data for 4U~0513-40.  We use all
data from proposals P10078, P30407, P40404, P50403, P60406, P70403,
P90402, P91402, P92403 and P93403, which include a total of 740
observations made between March 1996 and December 2008.  We take the
standard product light curves over the channels from 2-9 keV.  The
data have been filtered by the RXTE PCA team using the standard
criteria (a pointing offset less than 0.02 degrees from the target
position and an Earth elevation of at least 10 degrees).  All
proportional counter units with at least 95\% as much time turned on
as the counter with the largest ``on time'' are included.  The faint
source models are used for background subtraction in cases where the
count rate is less than 64 counts/sec/per PCU, which includes nearly
all of the observations of NGC 1851.  After background subtraction,
light curves are produced in counts/sec/PCU.

We then re-bin the standard products data from 16 seconds to 1024
seconds, using the lcurve task in XRONOS, so that there will typically
be 2-4 data points per observation.  This rebinning yields a
manageable total number of time bins in a single light curve
containing all data points for NGC~1851.  The count rates as a
function of time are presented in Figure 1, while in Figure 2, we zoom
in on a representative few months of the light curve, and in Figure 3,
we zoom in on an atypical portion of the light curve where the count
rate and amplitude of variability are smaller than typical.

In order to ensure that the relationship between the count rate and
the luminosity does not change strongly as a function of the spectral
state of the system, we have extracted spectra of two observations --
one which is among the highest count rate observations and one which
is among the lowest count rate observations.  The low count rate
observation, which has RXTE observation identification number
93403-01-13-02, was made on 19 October 2007.  Its X-ray spectrum is
well modelled by a $\Gamma=2.2$ power law, with the Galactic
absorption column density of $4.4\times10^{20}$ cm$^{-2}$, and with no
evidence for a strong thermal component.  The higher count rate
observation, which has RXTE observation identification number
93403-01-48-04, and which was made on 22 December 2008, is well fit by
Galactic absorption plus a 2.3 keV blackbody component which contains
most of the flux, and a $\Gamma=2.8$ power law tail.  The former
observation has a count rate of 7.2 counts per second per proportional
counter unit (PCU), while the latter has a count rate of 50.1 counts
per second per PCU, in both cases from 2-20 keV.  The former
observation has a flux of 9.8$\times10^{-11}$ ergs/sec/cm$^{2}$, while
the latter has a flux of $6.7\times10^{-10}$ ergs/sec/cm$^{2}$.  The
ratio of count rate to flux varies by only about 5\% between these two
observations, so, given that the count rate varies by a factor of more
than 10, the use of the count rate to trace the luminosity is well
justified.  In both cases, about 20\% of the source counts come from
9.0-20.0 keV, so the 2-9 keV count rates also are well representative
of the bolometric luminosity; the X-ray spectral differences are
largely reflected only in the high energy tail of the spectrum,
significant changes in the best fitting spectral model can occur with
only small changes in the bolometric luminosity to count rate ratio.

As an additional test of whether the count rate is a good tracer of
bolometric luminosity, we compute hardness ratios for 4U~0513-40 over
the full set of observations studied here, dividing the count rate
from 4-9 keV by the count rate from 2-4 keV.  These are plotted in
Figure 4.  The variations in hardness are smaller than the measurement
errors on the hardness.  We note that the observations made in 1996 do
have a different hardness ratio than the rest of the data, but we also
note that the RXTE gain was changed shortly after those data were
taken, moving the lower energy threshold for photon detection
substantially.\footnote{The RXTE gain epochs are documented at
http://heasarc.gsfc.nasa.gov/docs/xte/e-c\_table.html.}  The ratio of
Counts(9-20 keV)/Counts(4-9 keV) shown in Figure 5 is less susceptible
to this gain change and is consistent with being constant even across
the 1996-1998 time gap.

The X-ray flux of 4U~0513-40 varies from about 9$\times10^{-11}$
ergs/sec/cm$^2$ (about 20 times the RXTE confusion limit -- Jahoda et
al. 2006) to about $2\times10^{-9}$ ergs/sec/cm$^2$.  Assuming a
distance of 12 kpc (Alcaino 1976), this corresponds to a luminosity
variation from about $1.4\times10^{36}$ to $3\times10^{37}$ ergs/sec.
This is in agreement with the fact that the spectral state of this
system appears to be banana-like (i.e. occupying a particular region
in a colour-colour diagram that looks like a banana, and which is
characterized by being dominated by a quasi-thermal spectrum --
Hasinger \& van der Klis 1989) at its brightest, and island-like
(occupying a region in the X-ray colour-colour diagrams of Hasinger \&
van der Klis 1989 that looks like an island, and which is
characterized by relatively strong non-thermal emission) at its
faintest.  Therefore, the source makes a spectral state transition
near the bottom of its luminosity range, in the few per cent of
Eddington luminosity range where X-ray binaries normally make spectral
state transitions (Maccarone 2003).  The mean X-ray luminosity of
4U~0513-40 is about $3\times10^{36}$ ergs/sec and the mean accretion
rate is then about $3\times10^{16}$ g/s, assuming $L_X$$ =$$ 0.1$$
\dot{m} c^2$.  Theoretical tracks of mass accretion rate versus
orbital period predict accretion rates of $4\times10^{16}$ g/sec (to
within a factor of 2) for an ultracompact X-ray binary with an orbital
period of 17 minutes (Deloye \& Bildsten 2003).  The theoretical
uncertainty of a factor of 2 is determined primarily by the chemical
composition of the white dwarf (with higher mass transfer rates
expected from pure He white dwarfs than white dwarfs rich in carbon
and/or oxygen), but the initial temperature of the white dwarf also
can play a role, with hotter white dwarfs transferring mass more
quickly.

\begin{figure}
\psfig{figure=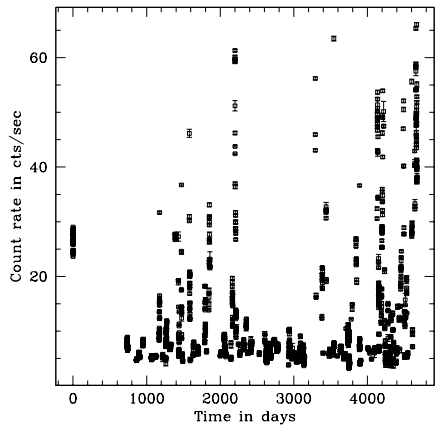,height=6 cm}
\caption{The full RXTE lightcurve for 4U~0513-40.  The x-axis gives the
time in units of days beginning from the start of the first
observation of the source, which occurred at 3:58:24 UT on 7 March
1996.  The y-axis shows the count rate in counts per second, from 2-9
keV.}
\end{figure}

\begin{figure}
\epsfysize=6cm \epsfbox{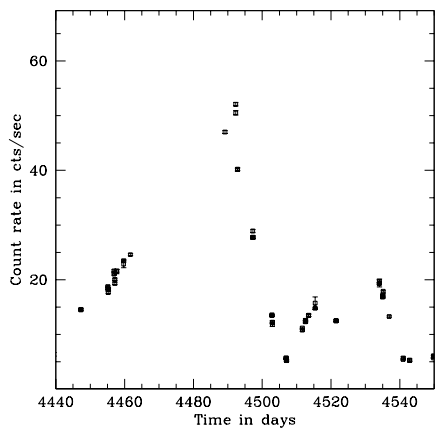}
\caption{The RXTE lightcurve for 4U~0513-40, from days 4450 to 4550
(i.e. starting from 13 May 2008).  The units on the axes are the same
as for figure 1.  The figure shows a typical large amplitude change in
the intensity of the source, which is of a factor of more than 10 in
count rate, and takes place on a timescale of a few weeks.}
\end{figure}

\begin{figure}
\epsfysize=6cm  \epsfbox{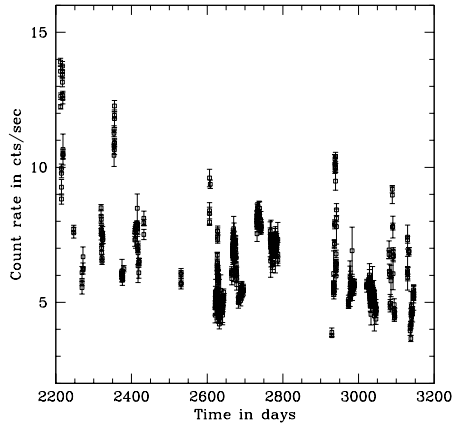}
\caption{The RXTE lightcurve for NGC~1851, from days 2200 to 3200
(i.e. from 27 March 2002 through 11 December 2004).  The units on the
axes are the same as for figure 1.  The amplitude of variations from
4U~0513-40 over the middle of this time span is considerably smaller
than over the spans before and after.}
\end{figure}

\begin{figure}
\epsfig{file=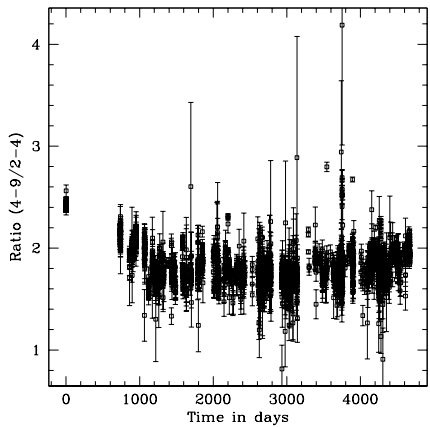,width=6 cm}
\caption{RXTE colours for NGC~1851.  The data presented are the ratio
of counts from 4-9 keV to the counts from 2-4 keV.  We note that while
it appears the hardness has changed substantially between the first
few points and the first few data points are from the short-lived
first RXTE gain epoch, while the remainder are all from epochs 3-5.
The largest change in the RXTE lower energy threshold came between
epochs 2 and 3.}
\end{figure}

\begin{figure}
\epsfig{file=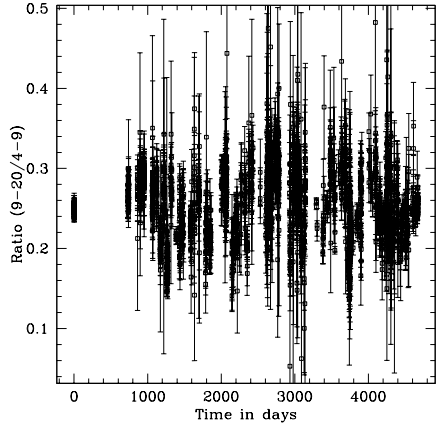,width=6 cm}
\caption{RXTE colours for NGC~1851.  The data presented are the ratio
of counts from 9-20 keV to the counts from 4-9 keV.}

\end{figure}

From the light curve plots, it is clear that 4U~0513-40 varies in
count rate by a factor of about 10.  One can then compare this
variation with those of more nearby persistent neutron star X-ray
binaries and find that this is a much larger amplitude of variability
than is typical.  The two most well-known Z-sources, Cygnus X-2 and
Scorpius X-1, vary by factors of about 3 in the dwell-by-dwell light
curve plots from the RXTE All-Sky Monitor
(http://xte.mit.edu/ASM\_lc.html, accessed 2 February 2010).  The two
systems which are classified as atoll sources in the catalog of Liu et
al. (2001), and which are confirmed to be ultracompact X-ray binaries,
4U~1820-303 and 4U~0614+091, both vary by factors of less than 3 over
the ASM light curves.  There is at least one neutron star source which
shows stronger variability without ever going into quiescence --
4U~1705-440 which varies by a factor of about 60 (see e.g. Homan et
al. 2009), but this system's orbital period is not known.  It thus
appears likely that the level of variability seen from 4U~0513-40 is
unusual for an ultracompact X-ray binary.  One the other hand, the
variations seen from 4U~0513-40 are considerably less than the factors
of $10^4$ or more in variability seen from bona fide transients.

The time sampling of 4U~0513-40 is insufficiently uniform for making a
reliable power spectrum without detailed simulations.  Nonetheless,
simple inspection of the light curve indicates strong variability on a
variety of timescales.  A series of sharp rises in the intensity are
seen from day 1400-2200 and from day 3400 to the end of the light
curve.  An epoch folding search for periodicities in the light curve
indicates marginal evidence for periodicity at about 88 days, but only
when rather long phase bins are used (1/8 or 1/16 of the period).
This is most likely due to this timescale being close to the spacings
of a few of the sharp increases in intensity, and is thus likely to be
spurious.  The period is not detected when finer bins are used for the
folding.  Indeed, it is clear from simple inspection of the light
curve between days 2400 and 3200, where the count rate never rises
above about 10 counts per second, that the light curve cannot be
described adequately as simply a periodic variation on a $\sim$90 day
timescale.  

We also looked for rapid variability within several individual
observations, but found that the source power spectra were generally
dominated by Poisson noise.  This was not surprising -- the source
spends most of its time at luminosities high enough that we expect it
to be in a ``banana'' state, where the integrated fractional rms
amplitude of variability should be only a few percent (see e.g. Berger
\& van der Klis 1998).  In the ``island states'', where the rms
fractional amplitude can approach 30\% (see again, for example, Berger
\& van der Klis 1998), the count rate will typically be only about 30
counts per second (for the rare observations where all PCUs are on),
and the background rate will be a substantial fraction of the total
count rate.  Additionally, the power is dominated by variability at
$\sim 10$ Hz in these states, so that more Poisson noise must be
integrated up to try to detect variability.  As a double-check on the
feasibility of detecting rapid variability, we can apply $n_\sigma=
\frac{1}{2} I r^2 \left({\frac{T}{\Delta \nu}}\right)^{\frac{1}{2}}$
from van der Klis (1989), which gives the signal to noise $n_\sigma$
in terms of the count rate $I$, fractional rms amplitude $r$, exposure
time $T$, and frequency width $\Delta \nu$.  Typical values in a
banana state are $\Delta \nu= 1$ Hz, $I=200$ counts/sec (assuming 5
PCUs), $r$=0.03 and $T$=3000 seconds, yielding a signal to noise of
2.5, while in an island state, $\Delta \nu= 10$ Hz, $I=30$ counts/sec
(assuming 5 PCUs), $r$=0.15 (30\% rms amplitude, but with only half
the counts from the source) and $T$=3000 seconds, yielding a
signal-to-noise of about 6 -- large enough to make a detection of
variability, but not large enough to use power spectral analysis to
help understand the source.

\subsection{Long timescale variability}
There are no strong luminosity flares observed between June 2002 and
March 2005 -- days 2300 through 3200 from the first observation in the
series.  This section of the light curve is plotted in Figure 3. The
mean count rate is $6.33\pm0.07$ counts/sec during the low flux epoch,
and $12.7\pm0.3$ during the rest of the light curve.  Comparing the
distribution of count rates during this epoch with the distribution
over the rest of the light curve using a Kolomogorov-Smirnov test
yields a probability of 1.6$\times10^{-58}$ that the same underlying
count rate distribution could produce these two epochs.  Therefore,
the data from the low count rate epoch could not have resulted from
uncorrelated fluctations in the count rates and hence reveal a real
physical change in the source.

\section{Discussion}

The variability from 4U~0513-40 has two important characteristics. The
first is that it is that the source manages to exhibit large amplitude
X-ray variability, while remaining persistently bright -- well above
the $\sim 10^{32}$ ergs/sec level at which crustal emission from the
neutron star, rather than accretion is the dominant source of X-rays
(e.g. Wijnands et al. 2001).  The second is that the time averaged
luminosity varies even if one averages over timescales as long as a
few years (see e.g. the lack of any strong flares between days 2400
and 3200).  Since such timescales are far longer than any reasonable
propagation timescale through a disc of such small radius, this
suggests real deviations of the mass transfer rate from that predicted
due to gravitational radiation, or at least some sort of disc
instability with a characteristic timescale far longer than the
viscous timescale.

Several classes of mass transferring binaries are at short enough
orbital periods that gravitational radiation is expected to dominate
over magnetic braking as the process responsible for the systems'
orbital evolution.  It has long been known that, in some cases where
the donor stars are non-degenerate, the mass transfer rates can exceed
the mass transfer rates predicted from gravitational radiation alone
(e.g. Osaki 1995).  However, the systems where strong deviations from
the expectations due to gravitational radiation have previously been
reported all had low mass main sequence donor stars, which are like to
be more susceptible to stellar atmospheric affects which could change
the accretion rate than are white dwarf stars.  Among both the
ultracompact X-ray binaries and the AM CVn stars (mass transferring
double white dwarfs), the persistent systems have shown mass transfer
rates consistent with the predictions from gravitational radiation
(Deloye \& Bildsten 2003; Roelofs et al. 2007), without any reported
evidence for large amplitude variability (albeit with small samples of
objects, most of which have not been monitored intensively).
Cataclysmic variables with main sequence/subgiant donor stars have in
many instances shown evidence for deviations of the accretion rates
from secular values on a range of timescales (e.g. Townsley \&
Gaensicke 2009).

\subsection{Possible causes}
The viscous timescale of an accretion disc is given by:
\begin{equation}
t_{visc} = 2\times10^6 {\rm sec} (\alpha/0.1)^{-4/5} \dot{M_{16}}^{-3/10} M_1^{1/4} R_{10}^{5/4}.
\end{equation}
Here, $\alpha$ is the dimensionless viscosity parameter,
$\dot{M_{16}}$ is the mass accretion rates in units of $10^{16}$
g/sec, $M_1$ is the mass of the accretor in solar masses, and $R_{10}$
is the radius from which the viscous timescale is being computed in
units of $10^{10}$ cm (Frank, King \& Raine 1995).  For 4U~0513-40,
using $\dot{m} \sim 3\times10^{17}$ g/sec as appropriate to the hot
state for the disk, $t_{visc}\sim 10 (\alpha/0.1)^{-4/5}$ days.  This
timescale agrees reasonably well with the durations of the increases
in luminosity seen from this system.

Given the reasonable (i.e. within factors of a few) agreement between
flaring timescales and the disc's viscous timescale, it is worth
considering whether we may be seeing a weak disc instability.
Considerable work has been done to estimate the accretion rates at
which discs are susceptible to ionization instabilities.  The lowest
observed luminosities from this system are at about $10^{36}$
ergs/sec, which corresponds to $1.1\times10^{16}$g/sec if
$L_X=0.1\dot{m}c^2$ as is typically assumed for neutron stars.  Menou
et al. (2002) presented the first calculation of the critical
accretion rates for hydrogen-poor discs, and found that for an orbital
period of 17 minutes, the critical accretion rate would be about
$1\times10^{16}$ g/s for discs with substantial carbon content, and
about $5\times10^{16}$ g/s for discs made of either pure helium or
pure oxygen.  Lasota et al. (2008) showed that irradation of the outer
disc by the X-rays produced in the inner disc, which were ignored in
the work of Menou et al. (2002) reduce the stability threshold for
$\dot{m}$ by a factor of about 4.  Thus while the accretion rate is
close to the threshhold for stability in the state-of-the-art
calculations of the disc instability model for hydrogen-poor gas, it
is unlikely that classical ionization instabilities are the proper
explanation for what we observe here.  Additionally, in at least some
of the flaring events, the rise times are slower than the decay times,
something which is not typical of ionization instability driven
outbursts.

The long timescale variations seen seem to imply that the actual mass
transfer rate from the white dwarf to the outer disk of the neutron
star is varying on a timescale of order a year.  As noted above, data
to date on mass transferring binaries with white dwarf donors has
found all are consistent with the mass transfer rates expected from
evolution driven purely by gravitational radiation, apart from the
known 179 day periodicity in the light curve of 4U~1820-30 (whose mean
luminosity agrees well with gravitational radiation predictions).  The
mean luminosity for 4U~0513-40 agrees, within the errors, with that
predicted based on the system's orbital period and the assumption that
the accretor is a $1.4M_\odot$ neutron star (see e.g. Deloye \&
Bildsten 2003 for predictions of X-ray luminosities based on realistic
models of white dwarf donors), but the aperiodic variations in the
accretion rate on long timescales suggest that real variations are
taking place in the accretion rate on these timescales.

A variety of mechanisms have been proposed for causing accretion rates
to vary from those expected based on the orbital and stellar
parameters in a binary system.  These include small orbital
eccentricities (e.g. Hut \& Paczynski 1984), perhaps due to the
presence of a third body in the system (Kozai 1962; Zdziarski et
al. 2007); tidal disc instabilities (e.g. Whitehurst 1988; Osaki
1995); star-spot blocking of the inner Lagrange point (e.g. Livio \&
Pringle 1994) and irradiation of the donor star leading to modulations
of $\dot{m}$ (e.g. Hameury, King \& Lasota 1986).  It is not clear
whether any of these mechanisms can explain either the short term
flaring or longer timescale variability we report here.  Also, the
lack of days to weeks variability in 4U~1820-30 and in the other
well-sampled ultracompact X-ray binary, 4U~1543-64 (Schultz 2003),
imply that whatever mechanism is at work for explaining the flaring
behaviour in 4U~0513-40 cannot generically apply to all ultracompact
X-ray binaries.

Most of the models above do not make specific predictions for how
variability would manifest itself in accreting neutron star systems.
However, a few statements can be made about the relative likelihoods
of the above mechanisms.  White dwarfs are not thought to have
convective regions, and hence the star spot model is not likely to be
relevant for these data.  The Kozai mechanism predicts strictly
periodic variations, rather than the aperiodic variability we see
here.  One could, in principle, involve a fourth body, in which case
chaotic behavior might be seen, but the probability of such a system
being formed is likely to be small.  Additionally, the outer bodies
would have to be quite faint, in order not to dominate the optical
flux from the system. 

The two other mechanisms, tidal instabilities and irradiation driven
mass transfer instabilities remain viable.  The chief criterion
required for tidal instabilities to be a possibility is to have large
mass ratio in the binary system (e.g. Whitehurst 1988; Osaki 1995), in
agreement with what is seen here.  Irradiation induced mass transfer
cycles involve nonlinear feedback and have not been studied in detail
in the context of systems with white dwarf donor stars, so it is
difficult to make any specific statements about their viability for
explaining what we have observed. Nonetheless, it is worth noting that
the original motivation for our investigation of the long term
variations of the X-ray luminosity in this system came from the
variations in the rms amplitude of the periodic signal in the
ultraviolet emission we used to find this system's orbital period
(Zurek et al. 2009).  We thus already found that it is likely that the
white dwarf in this system undergoes sufficient irradiation to lead to
a measurable difference in temperatures of its heated and unheated
faces, and that the extent to which the temperature varies across the
white dwarf's surface varies with time.  Irradiation induced mass
transfer cycles are then feasible, provided one can find a mechanism
by which they are seen prominently in 4U~0513-40, but are not seen
prominently in the brighter, tigher binary system 4U~1820-30.

\subsection{Implications for studies of extragalactic X-ray binaries}
The surprising results shown here may have important implications for
interpretations of data on extragalactic X-ray binaries.  Bildsten \&
Deloye (2004), for example, have suggested that most of the X-ray
binaries seen in elliptical galaxies are ultracompact X-ray binaries,
with the strongest piece of evidence in favor of this claim being the
similarity of the luminosity function predicted for ultracompact X-ray
binaries to that observed in the elliptical galaxies.  However, the
observations considered by Bildsten \& Deloye (2004) are predominantly
$\sim$ 40 kilosecond observations of galaxies as distances of
$\approx10-20$ Mpc.  A system with the same systems parameters
4U~0513-40, placed at such distances could be detectable only in its
brightest epochs.  However, shorter period systems could be expected
to be detected most of the time.  Finding large amplitude variability
to be common in ultracompact X-ray binaries would then imply that
snapshot X-ray luminosity functions of elliptical galaxies are not
reliable tracers of the time-averaged mass transfer rate distributions
even for ``simple'' X-ray binary systems like ultracompacts.

Additionally, large monitoring campaigns of elliptical galaxies have
recently begun to be undertaken such as the {\it Chandra} Very Large
Programs on the galaxies NGC~3379 and NGC~4278 (e.g. Brassington et
al. 2008,2009).  In these campaigns, the criterion for calling a
source a transient has been variability at the factor of 10 level.
However, the observations presented here show that variability at more
than the factor of 10 level is possible in sources that are expected
to be persistently accreting at high rates, and indeed, found to be
persistently accreting at high rates over timescales of at least a
decade.  As a result, if systems such as 4U~0513-40 are common, there
may be many objects classified as transients which are actually
persistent sources with large variability amplitudes.  A common
assumption is that the brightest (and hence shortest period)
ultracompact X-ray binaries will be steady sources, so the finding
that short period ultracompact X-ray binaries can be so strongly
variable should have important implications for such work.  At the
present time, it remains to be seen whether 4U~0513-40 is peculiar
among the persistent ultracompact X-ray binaries, or part of a
relatively common subset of sources; the combination of good sampling
in time with good sensitivity that could be provided by proposed
all-sky X-ray observatories such as LOBSTER (Fraser et al. 2002) would
be able to help answer this question.

\section{Conclusions}

The ultracompact X-ray binary 4U~0513-40 in the globular cluster
NGC~1851 shows large amplitude variability on a variety of timescales.
Of particular interest are the variations of factors of $\sim10$ on
timescales of several weeks and the variations in the luminosity when
time averaged over $\sim$ 1 year.  While it may be possible to explain
the $\sim$ weeks timescale variability through tidal instabilities in
the accretion disc, the latter almost certainly must be explained in
terms of modulation of the actual accretion rate.  Given that this
system's evolution is thought to be driven primarily by gravitational
radiation, and that the mean luminosity agrees well with that
scenario, the finding of significant deviations from a constant mass
transfer rate is interesting, and needs to be explained.

\section*{Acknowledgments}
We thank Phil Uttley for discussions about statistics.  We thank the
first referee Jean-Pierre Lasota for useful comments regarding
irradiated discs, and an anonymous second referee for several
constructive comments which improved this paper.  KSL and DRZ were
supported by NASA through grant GO-10184 from the Space Telescope
Science Institute, which is operated by AURA, Inc., under NASA grant
NAS5-26555.  This work makes use of results provided by the ASM/RXTE
teams at MIT and at the RXTE SOF and GOF at NASA's GSFC.

\label{lastpage}

\end{document}